\documentclass[conference]{IEEEtran}
\IEEEoverridecommandlockouts

\usepackage{cite}
\usepackage{amsmath,amssymb,amsfonts}
\usepackage{algorithmic}
\usepackage{graphicx}
\usepackage{textcomp}
\usepackage{xcolor}

\graphicspath{ {./resources/graphics/} }
\usepackage{listings}
\usepackage{multirow}
\usepackage{enumitem}
\usepackage{tcolorbox}
\usepackage{tabularx}
\usepackage{xcolor, colortbl}
\usepackage{subcaption}
\usepackage{xspace}
\usepackage{pgfplots}
\usepackage{booktabs}
\usepackage{pgfplotstable}

\newcommand{\tl}{25\xspace} 
\newcommand{\tfe}{7\xspace} 
\newcommand{\tp}{10\xspace} 
\newcommand{\tc}{153\xspace} 
\newcommand{\tf}{11\xspace} 

\definecolor{codebackground}{RGB}{255,255,255}
\definecolor{keywordcolor}{RGB}{0, 102, 204}
\definecolor{commentcolor}{RGB}{0, 153, 0}
\definecolor{stringcolor}{RGB}{204, 0, 0}

\lstdefinestyle{mystyle}{
    backgroundcolor=\color{codebackground},   
    commentstyle=\color{commentcolor}\itshape, 
    keywordstyle=\color{keywordcolor}\bfseries, 
    numberstyle=\tiny\color{gray},            
    stringstyle=\color{stringcolor},          
    basicstyle=\ttfamily\footnotesize,        
    breakatwhitespace=false,                   
    breaklines=true,                           
    captionpos=b,                              
    keepspaces=true,                           
    numbers=left,                              
    numbersep=5pt,                             
    showspaces=false,                          
    showstringspaces=false,                    
    showtabs=false,                            
    tabsize=2,                                 
    frame=topbot,                              
    rulecolor=\color{black},                   
    escapeinside={(*@}{@*)},                   
    morekeywords={*,...},                      
}

\AtBeginDocument{%
  }
    
\begin{document}

\title{Program Feature-based Fuzzing Benchmarking}

\author{\IEEEauthorblockN{Miao Miao}
\IEEEauthorblockA{\textit{Department of Computer Science} \\
\textit{University of Texas at Dallas}\\
Richardson, TX, USA \\
Email: mmiao@utdallas.edu}
}

\maketitle

\begin{abstract}
Fuzzing is a powerful software testing technique renowned for its effectiveness in identifying software vulnerabilities. 
Traditional fuzzing evaluations typically focus on overall fuzzer performance across a set of target programs, yet few benchmarks 
consider how fine-grained program features influence fuzzing effectiveness. 
To bridge this gap, we introduce a novel benchmark designed to generate programs with configurable, 
fine-grained program features to enhance fuzzing evaluations. We reviewed 25 recent grey-box fuzzing studies, 
extracting 7 program features related to control-flow and data-flow that can impact fuzzer performance. 
Using these features, we generated a benchmark consisting of 153 programs controlled by 10 fine-grained configurable parameters. 
We evaluated 11 popular fuzzers using this benchmark. The results indicate that fuzzer performance varies significantly based on 
the program features and their strengths, highlighting the importance of incorporating program characteristics into fuzzing 
evaluations.
\end{abstract}


\section{Research Problem and Motivation} 
The evaluation of fuzz testing is usually conducted on a set of target programs, focusing on the 
overall performance (e.g., bug finding capability and code coverage) after fuzzers run a preset period of time.
We observe that such evaluations often reveal that different fuzzers tend to favor specific programs. 
For instance, fuzzers’ performance often varies across different target programs in the 
evaluations that use FuzzBench~\cite{metzman2021fuzzbench}. 
One of the reasons for such variation lies in the design of any given fuzzer. 
For example, EcoFuzz’s advantage in reducing energy wastage and maximizing path coverage~\cite{yue2020ecofuzz} may become more 
pronounced as program complexity increases.
However, current evaluations do not consider program features or 
analyze performance deviations in relation to those features.
Therefore, the research community has yet to establish a link between fuzzing performance and program features; without this link,
it remains unknown if the hypotheses and claims made in these fuzzers hold, making it hard to assess and further improve them. 
To close this gap, we propose to develop a feature-based fuzzing benchmark that systematically controls the 
syntactic structure of programs. 
Our approach provides fine-grained, configurable parameters to construct benchmark programs that represent specific 
features of control-flow and data-flow complexity, offering deeper insights into fuzzing performance in terms of 
different program features.

\section{Background and Related Work}
There are several important fuzzing benchmarks and are widely used in fuzzing evaluation.
FuzzBench~\cite{fuzzbench} provides an infrastructure to evaluate fuzzers in terms of code coverage and 
vulnerability exposure. 
Magma~\cite{magma} provides a benchmark of 138 ported bugs in 9 open source programs 
along with the lightweight oracle (ground truth) that reports the bug when triggered.
There exist other benchmarks focus on vulnerabilities in the programs (e.g., FixReverter~\cite{fixreverter}, LAVA-M~\cite{lava}, 
and CGC~\cite{darpa-cgc}). 
These benchmarks focus on bugs or vulnerabilities in the programs but do not 
identify program properties that can influence the performance of the fuzzer. 
UNIFUZZ~\cite{li2021unifuzz} proposes a collection of pragmatic performance metrics to evaluate fuzzers from six complementary 
perspectives.  
GreenBench~\cite{ounjai2023green} focuses on energy consumption of fuzzing evaluations. 
Although these approaches bring new dimensions to understand fuzzer performance, they still overlook the influence of 
specific characteristics in target programs that can impact fuzzer effectiveness. 
Wolff et al.~\cite{wolff2022explainable} and Zhu et al.~\cite{zhu2019feature} evaluated fuzzers based on program properties. 
However, the program properties they proposed do not focus on the systematic generation of programs using 
configurable program features.

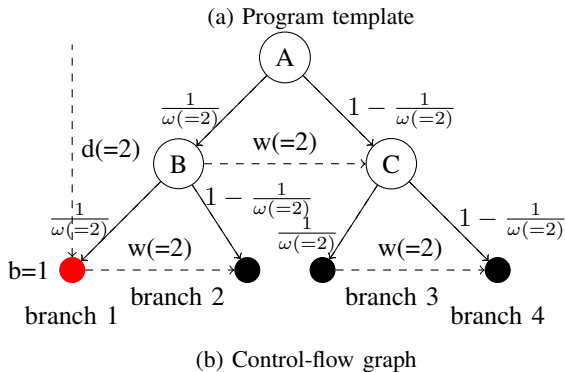
\begin{figure}[h]
    \centering
    \begin{minipage}{0.4\textwidth} 
        \begin{lstlisting}[language=C, escapechar=@, mathescape = true]
            
  void COMP_W2_D2_$\omega$2_B1(unsigned hash)
  {if (hash < 2) {  // A (nesting n=1)
       if (hash < 1) {  // B (nesting n=2)
           PRINTF("This is branch 1\n");
           ... // Inserting a bug here
       } else {
           PRINTF("This is branch 2\n");
       }
   } else {
       if (hash < 3) {  // C (nesting n=2)
           PRINTF("This is branch 3\n");
       } else {
           PRINTF("This is branch 4\n");
       ... // some code
  }
\end{lstlisting}
\subcaption{Program template}
    \end{minipage}
    \begin{minipage}{0.45\textwidth}
        \begin{tikzpicture}[node distance=2cm, auto]
  
            \node[draw, circle] (A) {A};
            \node[draw, circle, below left of=A] (B) {B};
            \node[draw, circle, below right of=A] (C) {C};
            \node[draw, circle, fill, below left of=B, red] (D) {};
            \node[left=5pt] at (D) {b=1};
            \node[below=10pt] at (D) {branch 1};
            \node[draw, circle, fill, below right of=B, xshift=-0.5cm] (E) {};
            \node[below=10pt, left=5pt] at (E) {branch 2};
            \node[draw, circle, fill, below left of=C, xshift=0.5cm] (F) {};
            \node[below=10pt, right=5pt] at (F) {branch 3};
            \node[draw, circle, fill, below right of=C] (G) {};
            \node[below=10pt] at (G) {branch 4};
          
            \path[->] (A) edge node {} (B)
                          edge node {} (C);
            \path[->] (B) edge node {} (D)
                          edge node {} (E);
            \path[->] (C) edge node {} (F)
                          edge node {} (G);
            
            \path[->, dashed] (B) edge node[above] {w(=2)} (C);
            \path[->, dashed] (D) edge node[above] {w(=2)} (E);
            \path[->, dashed] (F) edge node[above] {w(=2)} (G);
          
            \path[->, dashed] (A) edge node[left] {\( \frac{1}{\omega(=2)} \)} (B);
            \path[->, dashed] (A) edge node[right] {\( 1-\frac{1}{\omega(=2)} \)} (C);

            \path[->, dashed] (B) edge node[left] {\( \frac{1}{\omega(=2)} \)} (D);
            \path[->, dashed] (B) edge node[above=8pt, right=-8pt] {\( 1-\frac{1}{\omega(=2)} \)} (E);
            \path[->, dashed] (C) edge node[below=5pt, left=2] {\( \frac{1}{\omega(=2)} \)} (F);
            \path[->, dashed] (C) edge node[right] {\( 1-\frac{1}{\omega(=2)} \)} (G);

          
            \draw[<-, dashed] (D) -- ++(0,3) node[midway, right] {d(=2)};
          
          \end{tikzpicture}
          \subcaption{Control-flow graph}
    \end{minipage}
    \vspace{-6pt}
    \caption{Illustrative example of control-flow complexity parameters. \texttt{Width}, \texttt{Depth}, 
    \texttt{Weight}, and \texttt{BBranch} are denoted as \textit{w}, \textit{d}, \textit{$\omega$}, and 
    \textit{b} respectively.}
    \label{fig:controlflow}
\end{figure}


\section{Approach} 
\subsection{Feature Extraction}
We reviewed \tl grey-box fuzzing papers that are published within last three years, as well as the most cited 
fuzzers from earlier years, and summarized the common hypotheses or claims of improvements on 
fuzzing performance.
Note that It is not our goal to cover all published fuzzing papers. Instead, we reviewed these popular fuzzing 
papers as a representative set to extract important program features to construct the benchmark.
In total, we extracted \tfe program features from two aspects: \textit{control-flow complexity} and \textit{data-flow complexity}.
\subsubsection{Control-Flow Complexity (15 papers)}
Defined by four program features: \textit{number of conditional branches}, 
\textit{execution probability of conditional branches}, \textit{loops and recursions} and 
\textit{loops and recursions with data constraints}.

\subsubsection{Data-Flow Complexity (10 papers)}
Defined by three program features:
\textit{magic bytes}, \textit{checksum tests}, and \textit{nested magic bytes and checksum tests}.




\subsection{Benchmark Generation}
We generate synthetic programs emphasizing specific features with varying levels of strength to assess fuzzer 
performance. 
We adjust control- and data-flow complexity by stacking template blocks, and use fine-grained configurable 
parameters to control each feature's strength. 
We crafted \tp configurable parameters and generated a total of \tc programs, targeting \tfe distinct 
program features.

\subsubsection{Control-Flow Complexity.}
We define six parameters to manipulate the control-flow complexity of the programs.
\texttt{Width}, \texttt{Depth}, \texttt{Weight}, \texttt{BBranch}, are used to control 
the number of conditional branches and the execution probability of the buggy branch (as illustrated in Figure 
\ref{fig:controlflow}). 
\texttt{Iteration} and \texttt{Has\_Data\_Constraint}, are used to control the 
generation of programs with bugs reside in a deep loop or recursive call. 

\subsubsection{Data-Flow Complexity}
We use four parameters to define the data-flow complexity of the programs: \texttt{Start},
\texttt{Length}, \texttt{Depth}, and \texttt{Count}, which defines the starting index of the magic bytes, 
the number of involved magic characters, the nesting level of conditions, and the number of checksum tests, respectively.

\begin{table}[t]
    \small
    \centering
    \caption{\textbf{COMD}, \textbf{COMW}, and \textbf{COMWE} stand for Depth, Width, and Weight of 
    control-flow complexity. Spearman correlation (\textbf{corr}) and completion rate 
    (\textbf{comp}). RedQ stands for RedQueen, Mem-S and -H stand for two variants (Stack and Heap) of Memlock, 
    and Tort-B and -L stand for two variants (Basic Block and Loop) of TortoiseFuzz, 
    Hongg stands for Honggfuzz. Statically significant correlations are denoted with an asterisk (*).
    Weak correlations (between -0.3 and 0.3) with a 100\% completion rate are highlighted 
    in bold, and a hyphen (-) indicates unavailable correlations due to insufficient data.}
    \label{tab:corr_com}
    \begin{tabular}{lcccccc}
    \toprule
    \textbf{Fuzzer} & \multicolumn{2}{c}{\textbf{COMD}} & \multicolumn{2}{c}{\textbf{COMW}} & \multicolumn{2}{c}{\textbf{COMWE}} \\
    \cmidrule(lr){2-3} \cmidrule(lr){4-5} \cmidrule(lr){6-7}
     & \textbf{corr} & \textbf{comp} & \textbf{corr} & \textbf{comp} & \textbf{corr} & \textbf{comp} \\
    \midrule
    EcoFuzz & \textbf{0.287*} & 1.00 & \textbf{-0.024} & 1.00 & \textbf{-0.237*} & 1.00 \\
    MOpt & 0.662* & 1.00 & \textbf{0.106} & 1.00 & \textbf{-0.192*} & 1.00 \\
    AFLFast & 0.517* & 1.00 & \textbf{0.010} & 1.00 & -0.307* & 1.00 \\
    Fairfuzz & 0.513* & 1.00 & \textbf{0.141*} & 1.00 & -0.364* & 1.00 \\
    RedQ & 0.878* & 1.00 & - & 0.06 & -0.452* & 1.00 \\
    Laf-intel & 0.853* & 0.75 & 0.333* & 0.38 & \textbf{-0.291*} & 1.00 \\
    Mem-S & 0.534* & 1.00 & \textbf{0.154*} & 1.00 & \textbf{-0.228*} & 1.00 \\
    Mem-H & 0.500* & 1.00 & \textbf{0.177*} & 1.00 & \textbf{0.061} & 1.00 \\
    Tort-B & 0.839* & 1.00 & 0.253* & 0.94 & -0.474* & 1.00 \\
    Tort-L & 0.735* & 0.88 & 0.044 & 0.50 & -0.410* & 1.00 \\
    AFL & 0.640* & 1.00 & \textbf{0.255*} & 1.00 & -0.315* & 1.00 \\
    AFL++ & 0.894* & 1.00 & 0.872* & 1.00 & -0.507* & 1.00 \\
    Hongg & 0.366* & 1.00 & 0.013 & 0.94 & \textbf{-0.093} & 1.00 \\
    \bottomrule
    \end{tabular}
\end{table}

\section{Results and Contributions} 
We evaluate the performance of \tf fuzzers on our feature-based benchmark suite. 
We report the \emph{completion rate}, which calculates the successfully completed programs within the timeout 
to show how effectively each fuzzer supports specific feature parameters. 
We also calculate the \emph{Spearman's rank correlation coefficient} of each feature parameter and the fuzzing 
runtime to analyze the impact of the strength of each parameter on the performance of different fuzzers. 
Table \ref{tab:corr_com} shows the partial results of \textit{control-flow complexity} features parameters. 

We made several observations on different features. For example, \texttt{Depth} of control-flow 
complexity has a stronger impact on the fuzzing performance than \texttt{Width} of control-flow complexity. 
AFL++ is most sensitive to the increase of control-flow complexity, while EcoFuzz is the least sensitive. 
\emph{Our findings indicate that fuzzer performance varies significantly based on program features and their strength. 
Establishing a link between fuzzing performance and program features can help developers assess and improve these 
fuzzers for better effectiveness.}
Moving forward, we plan to perform static analysis to extract additional program features from real-world programs, 
and expand our benchmark to include features representing a broader range of real-world scenarios.

Overall, we made the following contributions in this work:
\begin{itemize}
	\item A literature review of \tl recent grey-box fuzzing papers to extract \tfe fine-grained program 
	features from their claimed improvements. 
	\item A feature-based fuzzing benchmark with 153 programs systematically generated using \tp configurable 
    parameters for the extracted program features.
	\item Evaluate \tf popular fuzzers on our feature-based benchmark to understand fuzzer behaviors and the 
    impact of each program parameter on their performance.
\end{itemize}


\section{Acknowledgment}
This work was partly supported by NSF grants CCF-2008905 and CCF-2047682.

\bibliographystyle{IEEEtran}
\bibliography{refs}

\begin{thebibliography}{10}
\providecommand{\url}[1]{#1}
\csname url@samestyle\endcsname
\providecommand{\newblock}{\relax}
\providecommand{\bibinfo}[2]{#2}
\providecommand{\BIBentrySTDinterwordspacing}{\spaceskip=0pt\relax}
\providecommand{\BIBentryALTinterwordstretchfactor}{4}
\providecommand{\BIBentryALTinterwordspacing}{\spaceskip=\fontdimen2\font plus
\BIBentryALTinterwordstretchfactor\fontdimen3\font minus
  \fontdimen4\font\relax}
\providecommand{\BIBforeignlanguage}[2]{{%
\expandafter\ifx\csname l@#1\endcsname\relax
\typeout{** WARNING: IEEEtran.bst: No hyphenation pattern has been}%
\typeout{** loaded for the language `#1'. Using the pattern for}%
\typeout{** the default language instead.}%
\else
\language=\csname l@#1\endcsname
\fi
#2}}
\providecommand{\BIBdecl}{\relax}
\BIBdecl

\bibitem{metzman2021fuzzbench}
J.~Metzman, L.~Szekeres, L.~Simon, R.~Sprabery, and A.~Arya, ``Fuzzbench: an
  open fuzzer benchmarking platform and service,'' in \emph{Proceedings of the
  29th ACM joint meeting on European software engineering conference and
  symposium on the foundations of software engineering}, 2021, pp. 1393--1403.

\bibitem{yue2020ecofuzz}
T.~Yue, P.~Wang, Y.~Tang, E.~Wang, B.~Yu, K.~Lu, and X.~Zhou,
  ``$\{$EcoFuzz$\}$: Adaptive $\{$Energy-Saving$\}$ greybox fuzzing as a
  variant of the adversarial $\{$Multi-Armed$\}$ bandit,'' in \emph{29th USENIX
  Security Symposium (USENIX Security 20)}, 2020, pp. 2307--2324.

\bibitem{fuzzbench}
\BIBentryALTinterwordspacing
J.~Metzman, L.~Szekeres, L.~Simon, R.~Sprabery, and A.~Arya, ``Fuzzbench: an
  open fuzzer benchmarking platform and service,'' in \emph{Proceedings of the
  29th ACM Joint Meeting on European Software Engineering Conference and
  Symposium on the Foundations of Software Engineering}, ser. ESEC/FSE
  2021.\hskip 1em plus 0.5em minus 0.4em\relax New York, NY, USA: Association
  for Computing Machinery, 2021, pp. 1393--1403. [Online]. Available:
  \url{https://doi.org/10.1145/3468264.3473932}
\BIBentrySTDinterwordspacing

\bibitem{magma}
\BIBentryALTinterwordspacing
A.~Hazimeh, A.~Herrera, and M.~Payer, ``Magma: A ground-truth fuzzing
  benchmark,'' \emph{Proc. ACM Meas. Anal. Comput. Syst.}, vol.~4, no.~3, Nov.
  2020. [Online]. Available: \url{https://doi.org/10.1145/3428334}
\BIBentrySTDinterwordspacing

\bibitem{fixreverter}
\BIBentryALTinterwordspacing
Z.~Zhang, Z.~Patterson, M.~Hicks, and S.~Wei, ``{FIXREVERTER}: A realistic bug
  injection methodology for benchmarking fuzz testing,'' in \emph{31st USENIX
  Security Symposium (USENIX Security 22)}.\hskip 1em plus 0.5em minus
  0.4em\relax Boston, MA: USENIX Association, Aug. 2022, pp. 3699--3715.
  [Online]. Available:
  \url{https://www.usenix.org/conference/usenixsecurity22/presentation/zhang-zenong}
\BIBentrySTDinterwordspacing

\bibitem{lava}
B.~Dolan-Gavitt, P.~Hulin, E.~Kirda, T.~Leek, A.~Mambretti, W.~Robertson,
  F.~Ulrich, and R.~Whelan, ``Lava: Large-scale automated vulnerability
  addition,'' in \emph{2016 IEEE Symposium on Security and Privacy (SP)}, 2016,
  pp. 110--121.

\bibitem{darpa-cgc}
D.~CGC, ``Darpa cyber grand challenge (cgc),''
  \url{https://github.com/CyberGrandChallenge/}, 2018.

\bibitem{li2021unifuzz}
Y.~Li, S.~Ji, Y.~Chen, S.~Liang, W.-H. Lee, Y.~Chen, C.~Lyu, C.~Wu, R.~Beyah,
  P.~Cheng \emph{et~al.}, ``$\{$UNIFUZZ$\}$: A holistic and pragmatic
  $\{$Metrics-Driven$\}$ platform for evaluating fuzzers,'' in \emph{30th
  USENIX Security Symposium (USENIX Security 21)}, 2021, pp. 2777--2794.

\bibitem{ounjai2023green}
J.~Ounjai, V.~W{\"u}stholz, and M.~Christakis, ``Green fuzzer benchmarking,''
  in \emph{Proceedings of the 32nd ACM SIGSOFT International Symposium on
  Software Testing and Analysis}, 2023, pp. 1396--1406.

\bibitem{wolff2022explainable}
D.~Wolff, M.~B{\"o}hme, and A.~Roychoudhury, ``Explainable fuzzer evaluation,''
  \emph{arXiv preprint arXiv:2212.09519}, 2022.

\bibitem{zhu2019feature}
X.~Zhu, X.~Feng, T.~Jiao, S.~Wen, Y.~Xiang, S.~Camtepe, and J.~Xue, ``A
  feature-oriented corpus for understanding, evaluating and improving fuzz
  testing,'' in \emph{Proceedings of the 2019 ACM Asia Conference on Computer
  and Communications Security}, 2019, pp. 658--663.

\end{thebibliography}

\end{document}